\documentclass[aps,prl,twocolumn,superscriptaddress,showpacs,amsmath]{revtex4}

\usepackage{dcolumn}
\usepackage{bm}
\usepackage{amsfonts}
\usepackage{graphicx,graphics}
\usepackage{amssymb}
\usepackage{rotate}
\usepackage{subfigure}
\usepackage{epsf}
\usepackage{epsfig}
\usepackage{amsmath}
\usepackage{dcolumn}
\usepackage{color}
\usepackage{hyperref}

\newcommand{\hto}[1]{Ho$_2$Ti$_2$O$_7${#1}}

\newcommand{\eg}{{\it e.g. }}
\newcommand{\cf}{{\it cf.~}}
\renewcommand{\=}{\,=\,}
\newcommand{\+}{\,+\,}
\renewcommand{\-}{\,-\,}

\begin{document}

\title{Topological Sector Fluctuations and Curie Law Crossover in Spin Ice}

\author{L. D. C. Jaubert}\email{l.jaubert1@physics.ox.ac.uk}
\affiliation{Laboratoire de Physique, \'Ecole Normale Sup\'erieure de Lyon, Universit\'e de Lyon, CNRS, 46 All\'ee d'Italie, 69364 Lyon Cedex 07, France.}
\affiliation{Max-Planck-Institut f\"ur Physik komplexer Systeme, 01187 Dresden, Germany.} 
\affiliation{Theoretical Physics, Oxford University, Wolfson College, Oxford, OX1 3NP, United Kingdom.} 
\affiliation{Okinawa Institute of Science and Technology, 12-22 Suzaki, Uruma, Okinawa, 904-2234, Japan}

\author{M. J. Harris}
\affiliation{School of Divinity, University of Edinburgh, New College, Mound Place, Edinburgh, EH1 2LX, UK}

\author{T. Fennell}
\affiliation{Paul Scherrer Institut, 5232 Villigen PSI, Switzerland}

\author{R. G. Melko}
\affiliation{Department of Physics and Astronomy, University of Waterloo, Ontario, N2L 3G1, Canada}
\affiliation{Perimeter Institute for Theoretical Physics, Waterloo, Ontario N2L 2Y5, Canada}

\author{S. T. Bramwell}
\affiliation{London Centre for Nanotechnology and Department of Physics and Astronomy, University College London, 17Ð19 Gordon Street, London WC1H 0AH, UK}

\author{P. C. W. Holdsworth}
\affiliation{Laboratoire de Physique, \'Ecole Normale Sup\'erieure de Lyon, Universit\'e de Lyon, CNRS, 46 All\'ee d'Italie, 69364 Lyon Cedex 07, France.}

\date{\today}

\begin{abstract}
At low temperatures, a spin ice enters a Coulomb phase - a state with algebraic correlations and topologically constrained spin configurations.  In \hto{}, we have observed experimentally that this process is accompanied by a non-standard temperature evolution of the wave vector dependent magnetic susceptibility, as measured by neutron scattering. Analytical and numerical approaches reveal signatures of a crossover between two Curie laws, one characterizing the high temperature paramagnetic regime, and the other the low temperature topologically constrained regime, which we call the \textit{spin liquid Curie law}. The theory is shown to be in excellent agreement with neutron scattering experiments. On a more general footing, $i)$ the existence of two Curie laws appears to be a general property of the emergent gauge field for a classical spin liquid, and $ii)$ sheds light on the experimental difficulty of measuring a precise Curie-Weiss temperature in frustrated materials; $iii)$ the mapping between gauge and spin degrees of freedom means that the susceptibility at finite wave vector can be used as a \textit{local} probe of fluctuations among topological sectors.
\end{abstract}


\maketitle

In spin liquid systems, where an absence of symmetry breaking leads to a disordered phase, system spanning correlations that are hidden from conventional probes can be manifest in topological properties~\cite{Wen2002}. These extended correlations typically requires measurement of non-local quantities, such as the entanglement entropy \cite{Levin2006,Kitaev2006}, which may not immediately be accessible by experiment.
The quest for spin-charge separation~\cite{Anderson1987} in models for high temperature superconductivity has led to extensive studies of model quantum spin liquids~\cite{Lee2008,Balents2010} and quantum critical phenomena~\cite{Senthil2004,Isakov2012}, where topological order can be identified.  However, the interest in topology is not limited to quantum systems, as the constraints inherent to geometrically frustrated magnets or classical dimer systems also lead to extremely rich many body behaviour.
``Topological order'' may also be defined heuristically in classical spin liquids with an absence of conventional order, by a lack of ergodicity between sectors in configuration space with different topological invariants (see \eg Ref.~\cite{Castelnovo2007}).
The most familiar invariant is the ``winding number'' which can be used to define $Z_2$ or integer topological invariants in a variety of 2D and 3D systems.  Interestingly, novel topological invariants unique to three-dimensional have recently been identified \cite{Ran2011}, which could have broad implications for classical spin liquids in higher dimensions, with certain underlying emergent gauge structures arising due to local constraints \cite{Freedman2011}.

\begin{figure}[ht]
\centering
\rotatebox{0}{\includegraphics[width=\linewidth]{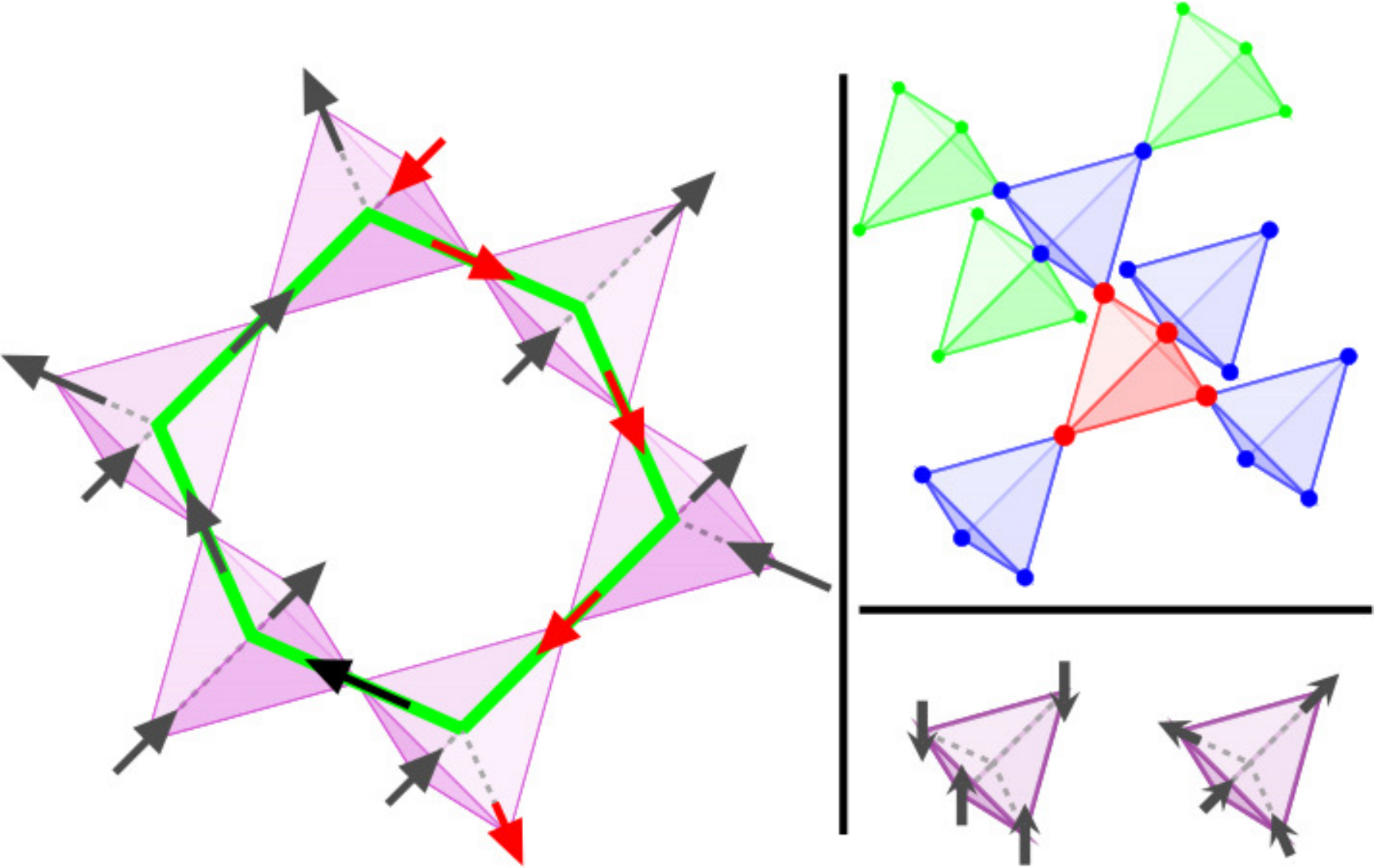}}
\begin{center}
\includegraphics[width=\linewidth]{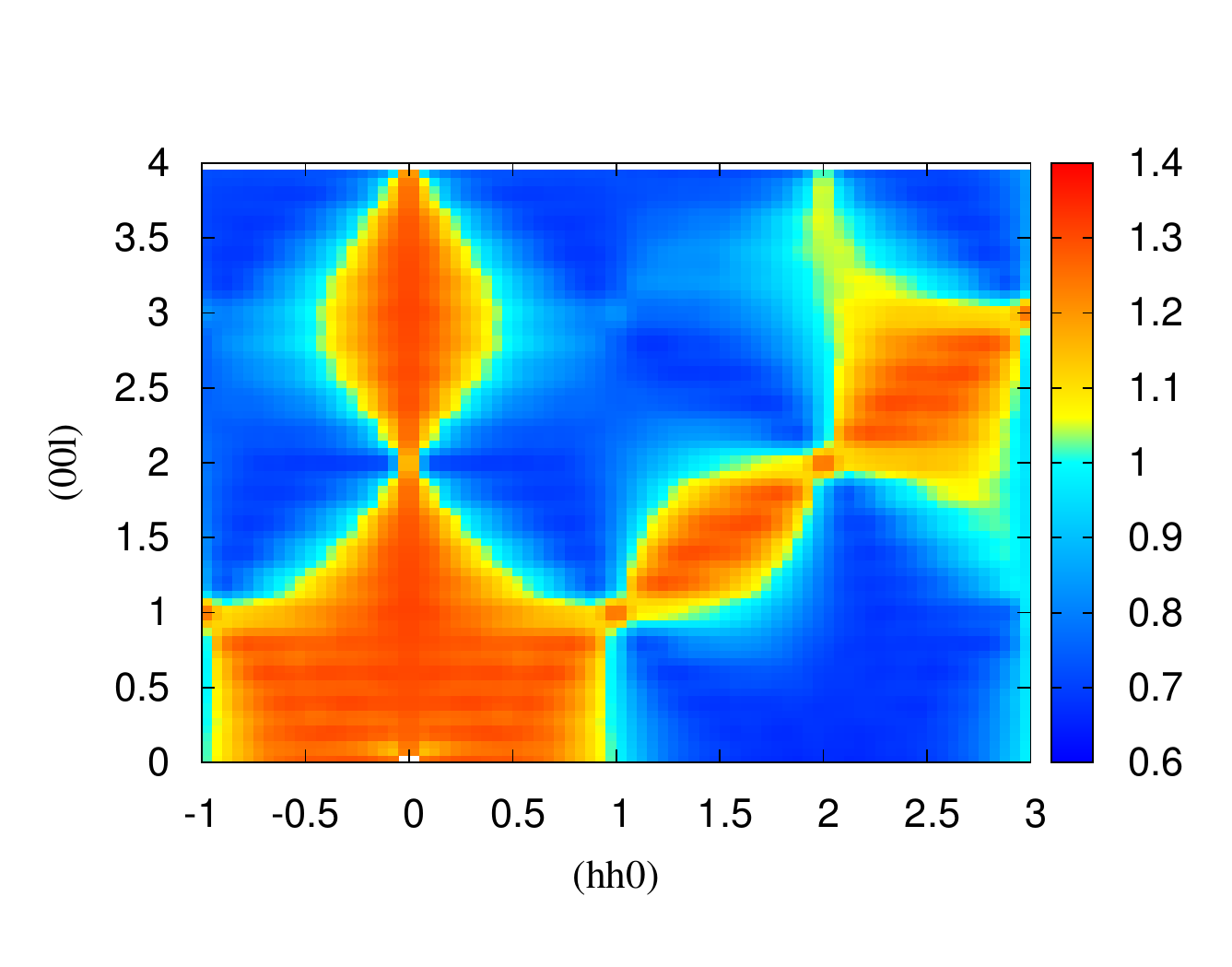}
\end{center}
\caption{\textit{Top right:} A portion of the pyrochlore lattice, made of corner-sharing tetrahedra.  All spins respect the ice-rules - ``two in two out''. The red spins form a string or worm in the negative direction, while the green hexagon represents a non-winding worm of 6 spins: flipping the former changes the topological sector, the latter is a fluctuation within a sector. \textit{Top right:} The Husimi tree construction for the pyrochlore lattice. \textit{Middle right:} Mapping between spins with easy-axis anisotropy respecting the symmetry constraints of the lattice (right), and the corresponding pseudo-spins (left). \textit{Bottom:} Scattering function $S(\mathbf{Q},T)$ of the NNSI model at $T=1$ K for a system of 4000 spins, simulating Ho$_{2}$Ti$_{2}$O$_{7}$ with $J_{\mathrm{eff}}=1.8$ K.  The pinch points are evident at $(0,0,2)$, $(1,1,1)$ and $(2,2,2)$.}
\label{fig:cell}
\end{figure}

The correlations induced by local constraints have been discussed in systems on the kagome lattice with both discrete~\cite{Macdonald2011} and continuous symmetry~\cite{Canals2001}, and for frustrated antiferromagnets with a pyrochlore structure~\cite{Isakov2004,Henley2005,Conlon2010}. In these theoretical models, the physics of the ``Coulomb phase'', with dipolar correlations showing up as pinch point singularities in reciprocal space~\cite{Zinkin1997,Fennell2007,Fennell2009}, is universally present.  Among systems showing Coulomb phase physics~\cite{Henley2010}, spin ice models~\cite{Bramwell1998} have proved particularly fruitful testing grounds for the collective behaviour associated with topological constraints as the 
associated spin ice materials are the only experimental magnetic systems for which sharp pinch points have been observed~\cite{Fennell2007,Fennell2009,Kadowaki2009,Chang2010}.  An extensive manifold of low energy states can be constructed by ensuring that a local constraint - the ice rule - is obeyed on every tetrahedron of the underlying pyrochlore lattice.  The ice rule requires that at low temperature, two spins point into and two out of every tetrahedron, as shown in Fig.~\ref{fig:cell}. This local divergence free condition also generates the long range dipolar correlations in the Coulomb phase, even within the confines of the simplest model with nearest neighbour spin interactions only.  

In this paper, we show that whilst a spin ice is topologically constrained, it is not topologically ordered as it is able to fluctuate between topological sectors.  We show how the susceptibility can be used as an indicator of these topological sector fluctuations (TSF), and present a detailed comparison of our results with both bulk magnetometry and  neutron scattering measurements on \hto{}~\cite{Bramwell2001a}. We compare the experimental data to analytical and numerical expressions of the susceptibility, where the emergence of TSF at low temperature appears as a crossover between two Curie laws at specific wave vectors $\mathbf{Q}$ in the structure factor $S(\mathbf{Q})$. This study was motivated in part by the experimental observation of an unusual  temperature dependence in the wave vector dependent susceptibility~\cite{Harris} (see Fig.~\ref{fig:SF:chiQ}), one that is fully explained by the theory presented here.

\paragraph{Topological sectors in spin ice:} Let us first characterize a model system, of size $L$, with periodic boundaries and only afterwards relate its properties to experimental observables.
We study the nearest neighbour spin ice model (NNSI)~\cite{Bramwell1998} with vector spins of unit length, $\vec S_i$ placed on the vertices of a pyrochlore lattice, constrained to lie along the body centred crystal field directions of the tetrahedra: $\vec S_i = \pm \vec d_i$ (see Fig.~\ref{fig:cell}) and with ferromagnetic exchange interaction $J > 0$.  The model maps onto Anderson's Ising antiferromagnet with exchange constant $J'=-J/3$~\cite{Moessner1998} through the definition of Ising pseudo-spin variables $\sigma_i = \vec S_i . \vec d_i$. However it is the ferromagnetic spin ice model that is physically realisable: the Ising pyrochlore antiferromagnet with a single easy axis cannot represent any real magnetic system as its Hamiltonian breaks the cubic lattice symmetry~\cite{Bramwell1998}. 
The ensemble of ground states satisfying the ice rules of two spins in and two out, or $\sum_{i=1,4} \sigma_i =0$, on each tetrahedron is a Coulomb phase~\cite{Isakov2004,Henley2005} and leads to the Pauling zero point entropy~\cite{Pauling1935,Ramirez1999} (see Fig.~\ref{fig:cell}). Here we consider cubic systems with $L^{3}$ unit cells and $N=16 L^{3}$ spins. Throughout the paper, as we have in mind comparisons with \hto, we take $J=1.8$ K, the value found to best parameterize thermodynamics measurements of this spin ice material~\cite{Bramwell2001a}.

A micro state of the Coulomb phase can be classified by a set of \textit{strings} of alternating out, in, out ... spins which  wind through the system along each of the cubic axes. That is, any spin $\vec S_{i}$ with a downward (upward) projection along a given cubic axis, always has at least one nearest neighbour above and below it with the same downward (upward) projection
(see Fig.~\ref{fig:cell}). Connecting these spins draws a map of strings spanning the system from top to bottom in the negative (positive) direction. We define an individual string, which does not necessarily close on itself, as an object spanning the system once along one cubic axis, so that each string is composed of $4L$ spins. 
Hence each microstate  has $n_{k}^{-}$ ($n_{k}^{+}$) strings spanning in the negative (positive) direction along cubic axis $\hat{k}$, such that 
the total number, $n_0=n_k^- + n_k^+= 4L^2$ equals the number of spins on a plane of the pyrochlore lattice perpendicular to $\hat{k}$. Each spin belongs to three strings threading along $\hat{x},\hat{y}$ and $\hat{z}$. Two strings threading through a tetrahedron in the same direction $\hat{k}$ are indistinguishable and can in fact be mapped onto the world lines of bosons living in the $d-1$ dimensional space perpendicular to the string direction\cite{Jaubert2008,Powell2008,Powell2011}. One can define a topological sector for each configuration through a  winding vector $\vec w=(w_{k}=n_{k}^{+}-n_{k}^{-})_{k=x,y,z}$, whose components are even integers taking values between $\pm 4 L^{2}$.

\begin{figure}[ht]
\includegraphics[scale=0.3]{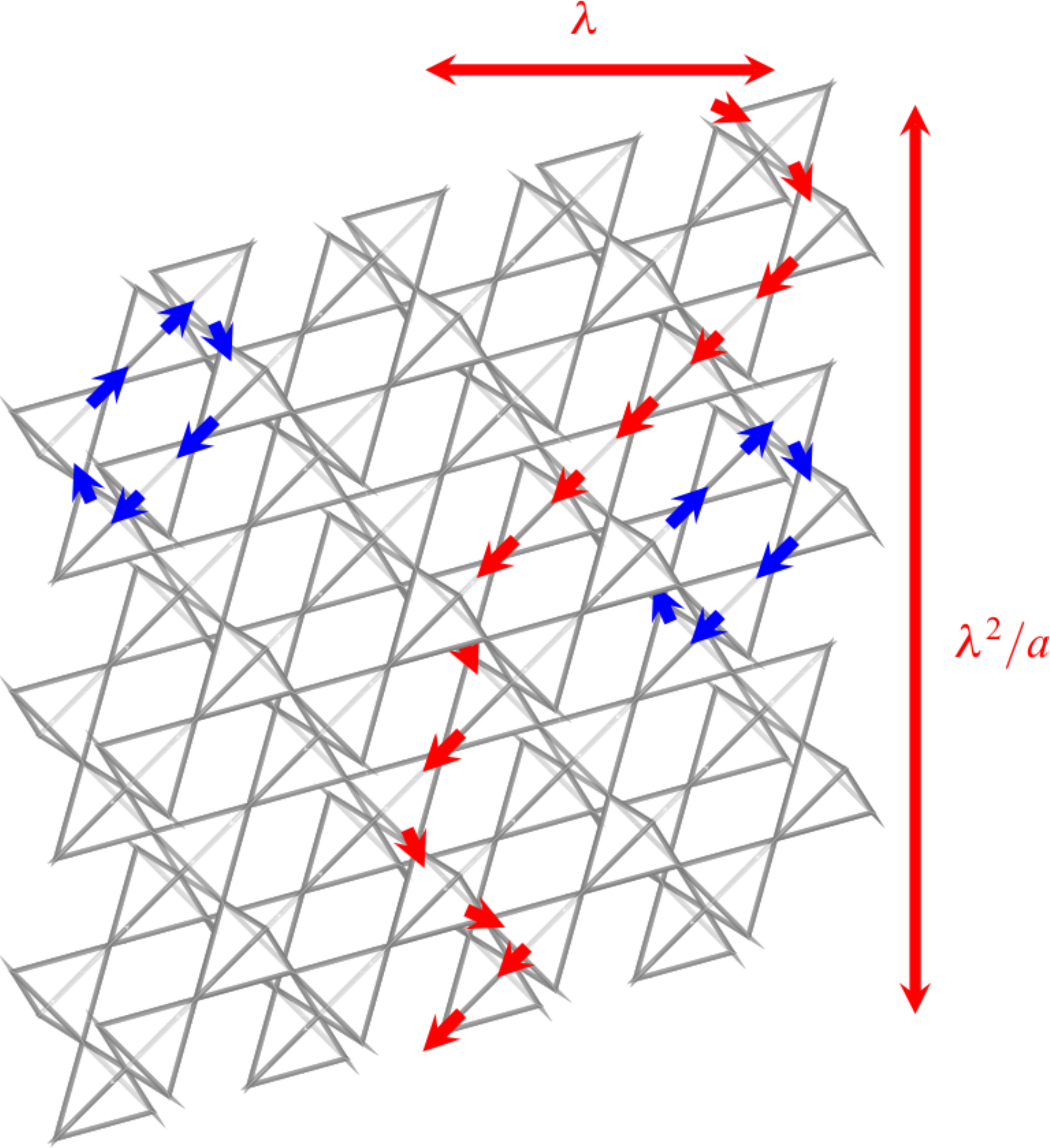}
\caption{Non-winding loop(blue) and string (red) which can be flipped by winding loops of this length. The winding loop excitation leads to a topological sector fluctuation. The (red) string illustrates the effective finite size system sampled by neutrons of wavelength $\lambda$.}
\label{fig:worms}
\end{figure}

Excitations within the Coulomb phase are non-local and limited to the flipping of a closed \textit{loop} of spins identified in two categories (see Fig.~\ref{fig:worms}): firstly a non-winding loop of spins which closes within the system. This kind of excitation moves the system between microstates of a given topological sector, re-arranging the string network without changing the winding number. Secondly, a winding loop which 
closes on itself after passing one or more times through the periodic boundaries. Each passage flips a string of spins and changes  the topological sector through a change in one component of the winding vector by two. 
We define a topologically ordered system as one that is restricted to a single topological sector.

In model systems the Coulomb phase space can be sampled directly in simulations by using a non-local \textit{worm} algorithm~\footnote{also called loop algorithm by Melko {\it et al}~\cite{Melko2004}.}. The worm is a virtual sequence of spins that burrows through the system until it closes on itself becoming a loop excitation that maintains the ice rules. In real materials, deconfined topological defects provide local dynamics~\cite{Jaubert2009b} and their creation and annihilation allows for the sampling of different constrained states. However, a finite concentration of such defects destroys the Coulomb phase above length scales fixed by their mean separation and the balance between maintaining ergodicity and imposing the constraints is a fine one.

The winding vector is a direct measure of the difference between upward and downward projections along each cubic axis, making it proportional to the magnetization: $\vec M = (4L/\sqrt{3}) \vec w$. 
Magnetic fluctuations of a system in the Coulomb phase are therefore a direct measure of topological sector fluctuations: non-winding loop excitations carry no magnetic moment, while winding ones carry a magnetic fingerprint of the change in topological sector. 
The phase space of constrained Pauling states making up the Coulomb phase is therefore compatible with an ensemble of gauge invariant topological sectors with $U(1)$ symmetry. Each sector is associated with an extensive subset of states with constant magnetization connected by non-winding loop excitations, while magnetic fluctuations correspond to a change in topological sector.
Susceptibility measurements could therefore be used as a diagnostic of both the crossover from the high temperature paramagnetic phase into the constrained phase and of the extent of topological sector fluctuations at low temperature. Applying a magnetic field along a cubic axis, $\hat{z}$, breaks the symmetry and the system approaches the maximum winding state, $\vec w=(0,0,4L^2)$ singularly via a Kasteleyn transition~\cite{Moessner2003a,Watson1999,Fennell2007,Jaubert2008,Powell2008,Morris2009}.  This is an example of a topological ordering transition, as the system enters a state with fixed sector and constant winding vector. The susceptibility diverges on the high temperature side but is strictly zero in the topologically ordered phase. In the rest of the paper we develop the notion of susceptibility as a diagnostic for TSF around $\vec w=0$, in zero field.

\paragraph{Analytics:} We calculate both the magnetic susceptibility and the pseudo-spin susceptibility on a Husimi tree of corner sharing tetrahedra, which preserves the coordinations of the pyrochlore lattice but neglects non-winding loops (see Fig.~\ref{fig:cell}). Although this is an approximate method, previous work~\cite{Jaubert2008} suggests that it should provide an excellent basis for describing magnetic fluctuations, as it allows system spanning strings of flipped spins and hence winding number fluctuations. The approach is reminiscent to the cluster variation method developed in~\cite{Yoshida2002}.
Spins on the $(n+1)^{th}$ shell (green in Fig.~\ref{fig:cell}) have three equivalent neighbours and one neighbour on the $n^{th}$ shell (blue in Fig.~\ref{fig:cell}). The total partition function is built up recursively by summing over the degrees of freedom of the spins from the $(n+1)^{th}$ shell, while holding the spin on the $n^{th}$ shell fixed, with an external field $h$ along the $z-$axis breaking the up down symmetry~\cite{Jaubert2009,Jaubert}. In order to neglect boundary effects, thermodynamic quantities such as the longitudinal susceptibility $\chi$, are extracted from the centre of the tree:
\begin{eqnarray}
\chi\equiv\left(\frac{\partial m_{z}}{\partial h}\right)_{h\rightarrow 0}\=
\frac{2\beta}{3}\;\frac{1+e^{2\beta J}}{2\+e^{2\beta J}\+e^{-6\beta J}},
\label{eq:SF:susc}
\end{eqnarray}
where $m_{z}=\langle\sum_i S_i^{z}/N\rangle$, $\langle...\rangle$ is a thermal average. Here and throughout our susceptibility has dimensions of inverse temperature. To compare with experiment it must be multiplied by $3C$ where $C$ is the normal Curie constant in the SI system : $C= \mu_0 \mu^2 N_{\rm Ho}/3 V \approx 4 $ K for \hto, where $\mu_0$ is the vacuum permeability, $\mu$ the magnetic moment operator for Ho$^{3+}$, $N_{\rm Ho}$ the number of Ho ions and $V$ the system volume. 

The asymptotic limits of $\chi(T)$~\cite{Isakov2004a,Ryzhkin2005} reveal a crossover between unconstrained and constrained, collective paramagnetic regimes, with the Curie constant scaled by a factor of two. 
\begin{eqnarray}
\chi(T\rightarrow \infty)\sim 1/3T \qquad \chi(T\rightarrow 0)\sim 2/3T.
\label{eq:SF:asymp}
\end{eqnarray}
The factor of $1/3$ at high temperature is a necessary property of a system with cubic space symmetry, familiar in the case of a Heisenberg paramagnet where an applied field couples to fluctuations in only one of the three Cartesian components of the magnetization. Although spin ice has local easy-axis anisotropy, it is isotopic in linear response once the full symmetry of the system is accounted for. This symmetry is a key property of spin ice and is related to the almost perfect screening of the long range interactions in the Coulomb phase of the dipolar spin ice model~\cite{Isakov2005}.
Further insight into this low temperature crossover can be gained from the pseudo-spin susceptibility (\cf figure~\ref{fig:cell})
\begin{eqnarray}
\chi_0\equiv\left(\frac{\partial m_0}{\partial h_0}\right)_{h=0}\=4\beta\;\frac{1+e^{-6\beta J}}{4\+6e^{2\beta J}\-2e^{-6\beta J}},
\label{eq:SF:susc0}
\end{eqnarray}
where $m_0=\sum_i  \sigma_i/N$ and where $h_0$ is parallel to the global pseudo-spin axis and thus conjugate to $m_0$. The asymptotic limits of $\chi_{0}$ are
\begin{eqnarray}
\chi_0(T\rightarrow \infty)\sim 1/T \qquad \chi_0(T\rightarrow 0^+)\sim {2e^{-2\beta J}\over{3T}}
\label{eq:SF:asymp0}
\end{eqnarray}
without the factor of $1/3$ as all pseudo-spins are parallel to the field. As the constraints are imposed, the pseudo-spin moment vanishes on each tetrahedron and $\chi_0$ falls to zero on the same temperature scale than the crossover of $\chi$ between the two Curie laws for the system with real spins (\cf figure~\ref{fig:SF:chiQ}.A). This reinforces our claim that this crossover can be used as a signal of the system entering the Coulomb phase, and as an indicator of topological sector fluctuations at low temperature.

Comparison with Monte Carlo simulations gives excellent agreement between analytics and numerics as shown in  Fig.~\ref{fig:SF:chiQ}.B, where we show both $T\chi$ and $T\chi_0$ from simulations of the NNSI using the worm algorithm. The data agree with the analytic prediction, within numerical error, over the entire range of temperature from $1000$ K to 0.5 K, at which point the system enters into the asymptotic regime characterized by equations (\ref{eq:SF:asymp}) and (\ref{eq:SF:asymp0}).
A closer look at $T\chi$ as $T\rightarrow 0$ taken from Ref.~\cite{Isakov2004a} gives an estimate of  $T\chi=0.66735 \pm 0.0003$, very close  (but not equal) to the $2/3$ predicted by the Husimi tree.  One can conclude that as the non-winding loops, which are absent in the tree calculation,  carry no magnetization this is of little consequence for magnetic fluctuations, allowing for an extremely accurate estimate of the magnetic fluctuations between different topological sectors.

\begin{figure}[ht]
\includegraphics[scale=0.4]{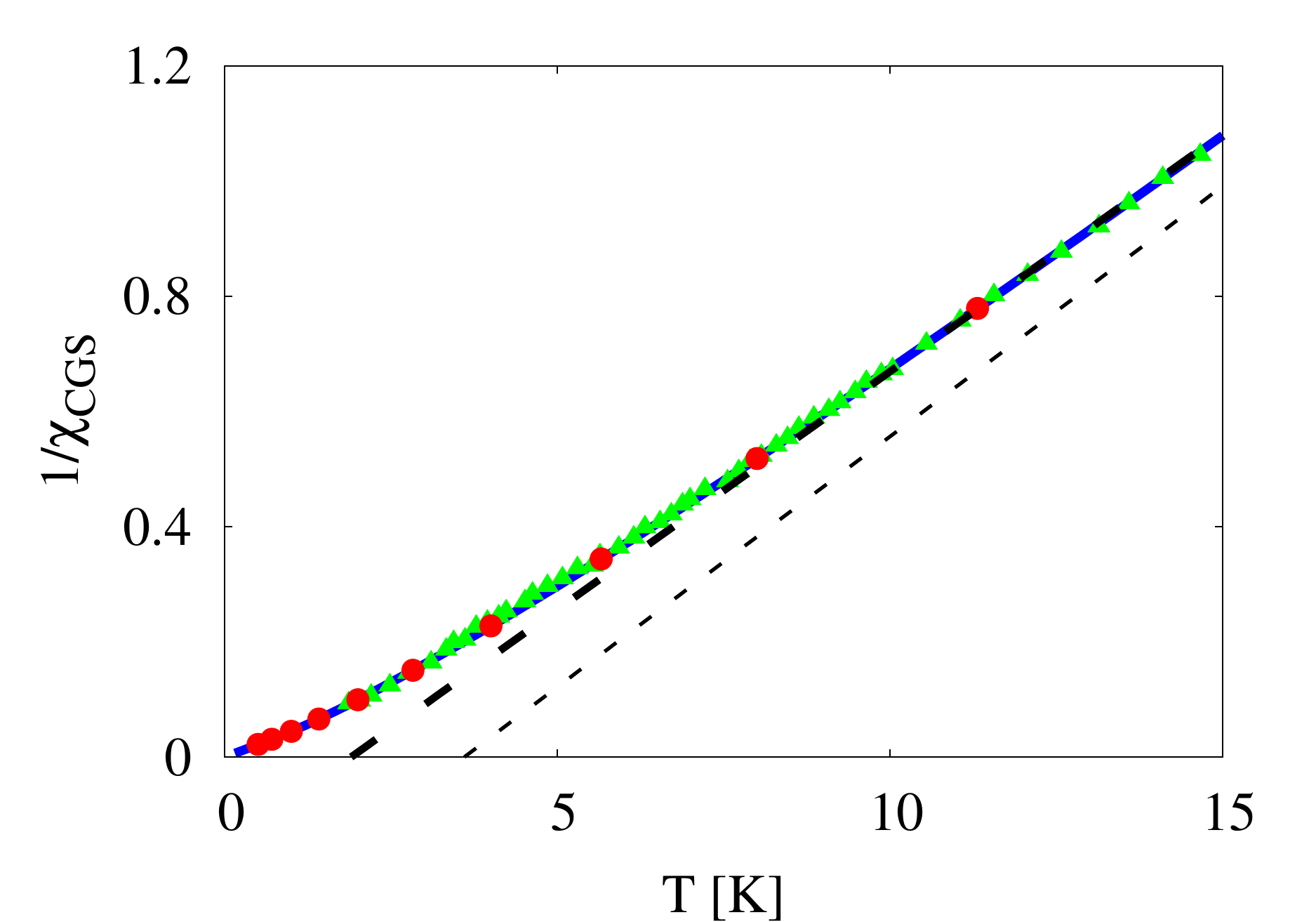}
\caption{$1/\chi$ vs. $T$ for bulk magnetometry measurements of a powder sample of \hto{} (green triangles), Monte Carlo simulation of the nearest neighbour spin ice (red dots) and analytical result from the Husimi tree, equation (\ref{eq:SF:susc}) (blue line). The dotted lines show Curie-Weiss laws for $\Theta_{CW}\equiv 2J=3.8 $ K (thin) and $\Theta_{CW}=1.9 $ K (thick). The susceptibility is expressed in CGS units cm$^{3}$.mol$^{-1}$.}
\label{fig:bulk2}
\end{figure}

\begin{figure*}
\includegraphics[width=\linewidth]{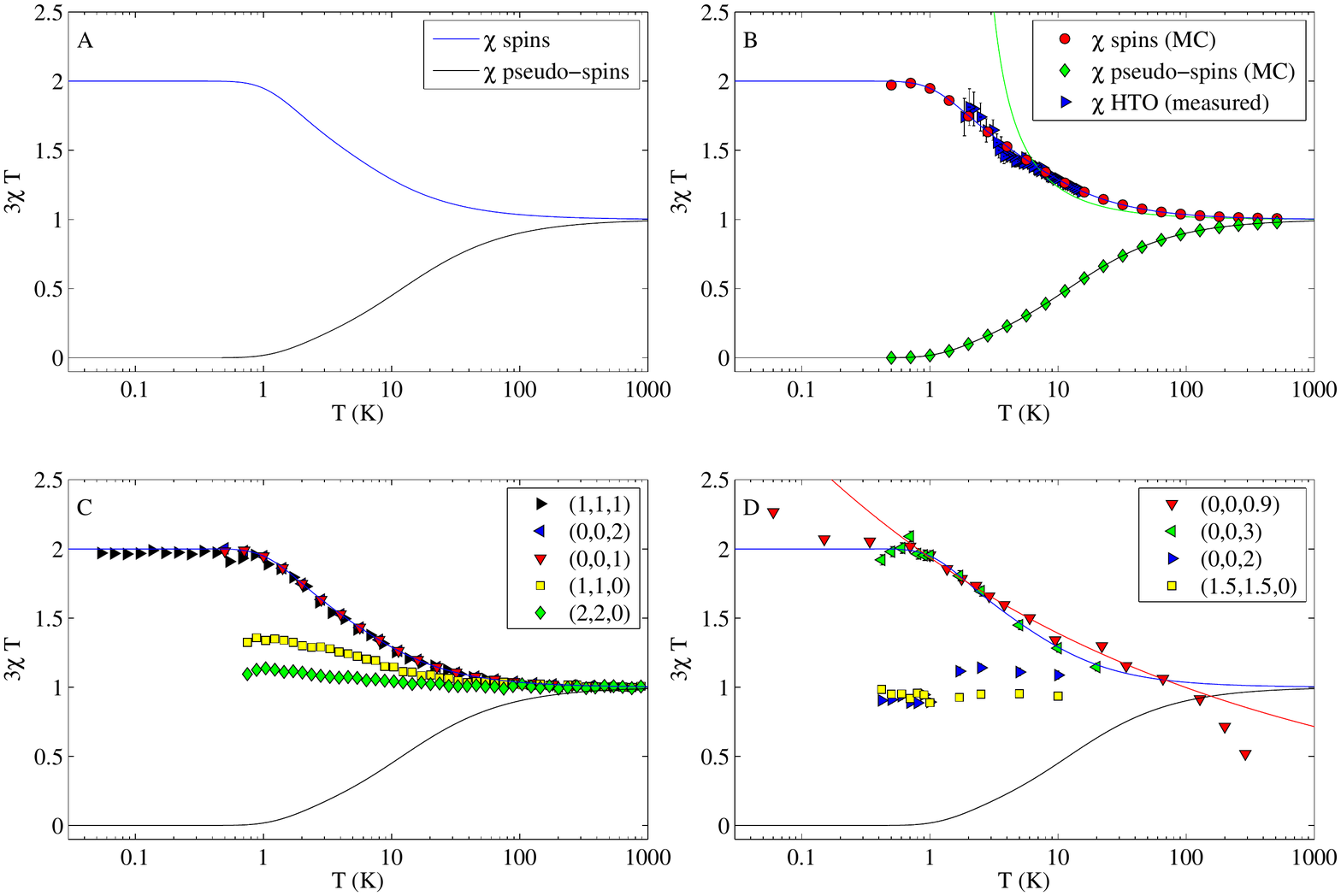}
\caption{\textit{A:} Temperature dependance of $3\chi T$ from Eq.~\ref{eq:SF:susc} with real spins (top curve) and $3 \chi_0 T$ From Eq.~(\ref{eq:SF:susc0}) with pseudo-spins (bottom curve) and with $J_{\mathrm{eff}}=1.8$ K.
 \textit{B:} Comparison of the susceptibility of the NNSI model obtained by Monte Carlo simulation with single spin flip dynamics compared with the analytical theory. The blue triangles are bulk susceptibility measurements fitted to theory with a factor of 96\% with respect to the expected value of the local magnetic moments of 10 $\mu_{B}$. We show for comparison the Curie-Weiss fit of figure~\ref{fig:bulk2} with $\Theta_{CW}=1.9$ K (green line). The bulk data have been extracted from figure F14.1 in~\cite{ISIS} and the error bars are due to digitalization.
 \textit{C:} Wave vector dependent susceptibility of the NNSI model at selected $\mathbf{Q}$ points, again obtained by Monte Carlo simulation, and compared with the analytical theory.  The data for $(0,0,2)$ and $(0,0,1)$ have been obtained using both single spin flip dynamics and a worm algorithm has been added for $(1,1,1)$ to circumvent the low temperature ergodicity loss~\cite{Melko2004}. The nearest neighbor coupling has been set to $J/k_B=1.8$ K, as estimated for \hto{}, within the nearest neighbour approximation~\cite{Bramwell2001a}.
 \textit{D:} Temperature dependance of $S(\mathbf{Q},T)$ measured experimentally (see text for details), compared with the analytical theory. The red line is a guide to the eye to illustrate that this crossover could be misinterpreted as a power law within a finite temperature window ($\chi\sim T^{-1.17}$ here). In each case the intensity is scaled to the value of the susceptibility estimated from the Husimi tree at a single temperature between $30$ and $100$ K. There are no fitting parameters for the temperature axis.}
\label{fig:SF:chiQ}
\end{figure*}

Moving from the model to real systems, there are no periodic boundaries, so no winding of loops but any finite window will have strings running through it much as in the periodic system. Although boundary effects  may change the string statistics \cite{Macdonald2011}, the same picture of strings and closed internal loops should hold, motivating detailed comparison between model systems and experiment. In Fig. \ref{fig:bulk2} we superimpose previously unpublished experimental data from susceptibility measurements for a powder sample of \hto{}  taken between $3$ and $15$K. The experimental moment was scaled to $96\%$ of its full value of $\mu=10\mu_B$ to get the best fit. 
Even when taking into account this scale factor (which is close to unity) the agreement between experiment,  theory and simulation is remarkably good, indicating that the bulk susceptibility does indeed approach the TSF regime as the temperature becomes of order $J$.

\paragraph{On the Curie-Weiss law:} Unfrustrated  ferro- or antiferromagnets order on a temperature scale set by the Curie-Weiss temperature, $|\Theta_{CW}|$, estimated from a high temperature expansion for the magnetic susceptibility
\begin{eqnarray}
(3 \chi T)^{-1} \propto 1 - \beta \Theta_{CW}.
\label{eq:CWlaw}
\end{eqnarray}
Expanding equation~(\ref{eq:SF:susc}) to order $\mathcal{O}(\beta)$ one finds  
$\Theta_{CW}=2J$ for the NNSI. A standard picture of frustrated compounds proposed by Ramirez~\cite{Ramirez1994} is that frustration will hinder ordering down to a lower temperature $T^{*} \ll \Theta_{CW}$, with  $T^{*}\rightarrow 0$ for a spin liquid, or cooperative paramagnet. This is the case for spin ice, although unlike an antiferromagnetic spin liquid, the susceptibility approaches what we call the \textit{spin liquid Curie law}, characteristic of topological sector fluctuations, rather than a constant value, as $T$ approaches zero. The crossover occurs over a very wide range of temperature, so that while the TSF regime is reached around $1$ K, the paramagnetic Curie law is only reached above $100$ K.
This is extremely important for comparison with experiment, as $100$ K is far outside the spin ice temperature range for real materials making an estimate of  $\Theta_{CW}$ by traditional methods a difficult task. For example,
putting $J=1.8 $ K, the canonical value  \hto{} gives $\Theta_{CW}=3.8 $ K, which is noticeably bigger than the estimate of $\Theta_{CW}=1.9 \pm 0.1 $ K from the bulk magnetometry measurements for \hto{} shown in Fig. \ref{fig:bulk2} \cite{Bramwell2001a}. The reason is that the rare-earth single ion Ho$^{3+}$ starts to lose its Ising nature and becomes more and more Heisenberg-like with easy-axis anisotropy above $\sim 30$K ~\cite{Clancy2009}. Although well above the scale set by $J$, it is still in the crossover region and far from the high temperature limit. In this crossover region, the effective measured value of the Curie-Weiss temperature thus depends on the temperature window used for the fit. Of course this does not prevent an estimate of the Curie-Weiss temperature, or the use of Ramirez's criterion mentioned above.
We illustrate this in Fig (\ref{fig:bulk2}) where we plot $1/\chi$ against temperature for the bulk susceptibility data together with theory and simulation. The data, now scaled by an effective moment which is $96\%$ of the full moment, compares extremely well with a Curie-Weiss law with $\Theta_{CW}\sim 1.9$ K, but lies far from the true Curie-Weiss law with $\Theta_{CW}=2J=3.8 $ K.

\paragraph{Simulations of the scattering function $S(\mathbf{Q},T)$:}
The scattering function $S(\mathbf{Q},T)$ measured from diffuse unpolarized neutron scattering intensity in the static approximation is defined as%
\begin{eqnarray}
S(\mathbf{Q},T)\equiv\left\langle\left|\sum_{i=1}^{N}\mathbf{S}_{i\,\perp}\,e^{i\mathbf{Q}\cdot\mathbf{r}_{i}}\right|^{2}\right\rangle,
\label{eq:SF:SQdef}
\end{eqnarray}
where $\mathbf{S}_{i\,\perp}$ is the component of a spin at $\mathbf{r}_{i}$ orthogonal to the scattering vector $\mathbf{Q}(q_x,q_y,q_z)$, and where $\langle...\rangle$ represents a thermal average at temperature $T$. Being the Fourier transform of the thermally averaged two-spin correlation function, $S(\mathbf{Q},T)$ is related to the wave vector dependent susceptibility, and in certain instances becomes an explicit function of the bulk susceptibility $\chi(T)$. For example in the paramagnetic phase, equation~(\ref{eq:SF:SQdef}) becomes
\begin{eqnarray}
S(\mathbf{Q},T\rightarrow\infty)=\left<\sum_{i=1}^{N}\;\left(S_{i\,x\prime}^{2}\+S_{i,y\prime}^{2}\right)\right>={2N\over{3}},
\label{eq:SQTinf}
\end{eqnarray}
where $x\prime$ and $y\prime$ are the axis of the plane orthogonal to $\mathbf{Q}$. Hence, since $\chi=1/3T$ in this regime, $S(\mathbf{Q},T) = 2 NT\;\chi(T)=2N/3$ for all $\mathbf{Q}$ as $T\rightarrow \infty$: that is, while $\chi$ is coupled to only \textit{one} spin component parallel to the field, $S(\mathbf{Q})$ is coupled to \textit{two} components orthogonal to the wave vector $\mathbf{Q}$~\footnote{This factor of two is unrelated to the apparition of magnetic correlations in the Coulomb phase, as in equation~(\ref{eq:SF:asymp})}.

As one moves into the Coulomb phase and correlations build up, $S(\mathbf{Q},T)$ develops a strong $\mathbf{Q}$ dependence with, in particular, the appearance of the pinch points \cite{Youngblood1981,Zinkin1997,Henley2005} that are characteristic of the local divergence free constraint imposed by the ice rules. A map of $S(\mathbf{Q},T)$ generated from equation (\ref{eq:SF:SQdef}),  in the $(h,h,l)$ plane of reciprocal space, for the NNSI as the Coulomb phase is approached, is shown in Fig.~(\ref{fig:cell}). The wave vectors are in units of $2\pi/a$, where $a$ is the side of a 16 site cubic unit cell. The pinch points, narrow regions of intense diffuse scattering can be seen at the reciprocal lattice vectors $(0,0,2)$; $(1,1,1)$; and $(2,2,2)$, that is, at Brillouin zone centres for the face centred cubic lattice of the pyrochlore structure. Near the  $(0,0,2)$ pinch point the scattering is expected to take the form
\begin{equation}
S(\mathbf{Q},T)= C(T){ {\tilde{q}_z^2 + \xi_{\mathrm{ice}}^{-2}(T)}\over{\tilde{q}_z^2+\tilde{q}_x^2+\tilde{q}_y^2+\xi_{\mathrm{ice}}^{-2}(T)}},
\label{spinch}
\end{equation}
where $\tilde{q}_x=q_x,\tilde{q}_y=q_y,\tilde{q}_z=q_z-2\times (2\pi/a)$ and where, following Youngblood and Axe~\cite{Youngblood1981},  $\xi_{\mathrm{ice}}(T)$ is a  coherence length for the Coulomb phase~\cite{Fennell2009,Henley2005}. 
The amplitude $C(T)$, which is the value of $S(\mathbf{Q})$ as one traverses the singular point at $\tilde{q}_x=\tilde{q}_y=0$, comes from transverse magnetic fluctuations: spin fluctuations in the plane perpendicular to the wave vector $\tilde{q}_z$ that describes them. It is therefore coupled to two spin components and so should constitute two contributions from the bulk susceptibility $\chi$. This expression predicts a ridge of intense diffuse scattering of constant amplitude along the cubic axis ($\tilde{q}_{x}=\tilde{q}_{y}=0$), whose width is limited by $\xi_{\mathrm{ice}}^{-2}$ at the pinch point:
$S(\mathbf{Q}=(0,0,l),T)=C(T)=2NT\chi$ which scales as $4N/3$ as $T$ goes to zero, while at higher temperature, as $\xi_{\mathrm{ice}}(T)$ becomes microscopic, the pinch point broadens and one crosses back to isotropic homogeneous paramagnetic scattering, with $S(\mathbf{Q},T)$ independent of $\mathbf{Q}$ and scaling as $\sim 2N/3$. As a summary, we expect
\begin{eqnarray}
S(\mathbf{Q},T)\=2 NT \;\chi(T)\;
\begin{cases}
\forall \mathbf{Q}\quad T\rightarrow \infty\\
\forall T \quad \vec{\tilde {q}}=0
\end{cases}
\label{eq:SQsummary}
\end{eqnarray}
In Fig.~\ref{fig:SF:chiQ}.C we show simulation results for the NNSI for $3S(\mathbf{Q},T)/2N$ as a function of temperature for different values of  $\mathbf{Q}$. For $\mathbf{Q}$ along $(0,0,l)$ the simulation results confirm the above scenario to a high degree of accuracy, as results for both $(0,0,1)$ and $(0,0,2)$ follow the theoretical expression given by the tree calculation within numerical precision between $T=0.1$ K and $T=1000$ K. We expect this argument to hold at any pinch point, as confirmed for (1,1,1) in Fig.~\ref{fig:SF:chiQ}.C. Away from the line of high symmetry $(0,0,l)$ and from pinch points, the scattering intensity fails to develop as the constraints are imposed, but remains larger than the high temperature asymptote.  As seen from the data at $(1,1,0)$ and $(2,2,0)$, the topological sector information is not contained in these projections, or at least in a less straightforward way.

\paragraph{Pinch point scattering:} Although Youngblood and Axe predict constant amplitude along $(0,0,l)$ it is perhaps surprising to find that this long wavelength expression holds all the way from the zone centre (pinch point) to the zone boundary. The $\tilde{\mathbf{q}} = (\tilde{q}_x,\tilde{q}_y,\tilde{q}_z)$ independence is a consequence of the collective paramagnetism yielding diffuse, rather than either Bragg or critical scattering: at finite $\mathbf{\tilde{q}}$, away from the zone center along the cubic axis $(0,0,l)$, one observes topological sector fluctuations in an effective system of reduced size $\sim 2\pi/|\tilde{\mathbf{q}}|$, which has only small finite size corrections to those in the thermodynamic limit. 

Specifically, for $\mathbf{q}$ along $\hat{z}$, we are interested in fluctuations in $M_x\propto w_{x}$ and $M_y\propto w_{y}$. The strings  of alternating ``out-in-out-in ...'' spins along a given cubic axis behave as random walkers in the plane perpendicular~\cite{Bhattacharjee1983,Jaubert2011}, so that those oriented along the $\hat y$ axis have a ballistic trajectory in this direction,  but make a diffusive random walk in the $(\hat x - \hat z)$ plane. For $\mathbf{q}$ along $\hat{z}$, string correlations will therefore be lost  when the extension of the string in the $\hat z$ direction exceeds $\lambda= 2\pi/|\mathbf{q}|$.
If the number of steps along the $\hat y$ axis is $\tilde{\ell}_y$, the diffusive orthogonal extension in the $(\hat x - \hat z)$ plane is $\sim \sqrt{\tilde{\ell}_y}$. 
For a string oriented along the $\hat y$ axis, each tetrahedron on its way provides 2 possible paths, alternatively along the [101] and [10$\bar{1}$] axes. Hence such string makes an almost isotropic random walk step in the $(\hat x - \hat z)$ plane after spanning 2 tetrahedra, along $\pm \hat x$ or $\pm \hat z$ approximately. Because there are 4 tetrahedra in a cubic cell, a step length is $a/2$.
Fixing the perpendicular extension $(a/2)\sqrt{\tilde{\ell}_y}=\lambda$ gives
$\tilde{\ell}_y \sim (2 \lambda/a)^2$, a number that is always greater than unity, even at the zone boundary, where $\lambda=a$ and $\tilde{\ell}_y$ spans two cubic cells. Hence, even at the zone boundary we are observing strings for a system of large enough effective size to be essentially in the asymptotic regime where one can observe topological sector fluctuations. We are not comparing with a system with periodic boundaries; the soft effective boundaries provided by finite scattering wave vector appear to give similar results, allowing constant scattering amplitude along the entire $(0,0,l)$ ridge.

To quantify this correlation, we can rewrite the susceptibility as
\begin{eqnarray}
3 \chi T&=&\frac{1}{N}\;\sum_{i,j} \left(\langle \mathbf{S}_{i}\cdot\mathbf{S}_{j}\rangle\-\langle \mathbf{S}_{i}\rangle\langle\mathbf{S}_{j}\rangle\right)\nonumber
\=1\+\sum_{i\neq0}\langle \mathbf{S}_{i}\cdot\mathbf{S}_{0}\rangle.
\label{eq:chithesis}
\end{eqnarray}
Obviously we don't expect any $\mathbf{q}$ dependence in the infinite temperature limit. In the Coulomb phase on the Husimi tree lattice, the total correlation between $i^{th}$ nearest neighbors is $\langle \mathbf{S}_{i}\cdot\mathbf{S}_{0}\rangle\=2/3^{i}$~\cite{Jaubert2009,Jaubert}. $\chi=2/3T$ is recovered upon integration of all correlations, and we can estimate that most of the correlations (98 \%) are already taken care by the first three nearest neighbors, i.e within a sphere of radius $r \lesssim a$.

In pictorial terms (see Fig. (\ref{fig:worms}), a neutron of wavelength $\lambda$ will not see closed loops on that scale, as they do not change the two point correlation function $g(r=\lambda)$, or magnetic moment calculated over an area incorporating the loop, but they will detect fluctuations on a larger scale ($r >\lambda$), which appear as fluctuations of strings as a result of the anisotropic scaling of the string trajectory.  Hence the strings should give essentially uniform scattering right up to the zone boundary.

\paragraph{Neutron scattering experiments on \hto{}:}
We now compare our theoretical and numerical findings with experimental measurements of $S({\bf  Q})$. In Fig.~\ref{fig:SF:chiQ}.D we show $S(\mathbf{Q})$ for several $\mathbf{Q}$ values in the $(h,h,l)$ plane from two different single crystal neutron scattering experiments of  \hto{} (that used different samples). Both experiments were performed on the IN14 triple axis spectrometer at the ILL, Grenoble. In the first experiment $S({\bf Q})$ was extracted from the measured unpolarized neutron scattering cross section~\cite{Bramwell2001}, while in the second it was derived by combining nonequivalent components of the tensor $S({\bf Q})$ measured by polarization analysis~\cite{Fennell2009}. In each case the intensity is scaled to the value of the susceptibility estimated from the Husimi tree at a single temperature between $30$ and $100$ K. There are no fitting parameters for the temperature axis.
One can immediately see that the total intensity at, or near the zone boundaries, $(0,0,0.9)$ and $(0,0,3)$ is in remarkably good agreement with theory and simulation, and shows the clear signature of a crossover between paramagnetic fluctuations and topological sector fluctuations. However, things are very different at the pinch point, $(0,0,2)$. After an initial increase above the Curie law below $30$ K, the value of the scattering intensity stagnates and even decreases as the temperature dips below $3$ K. This failure to follow the predictions of the NNSI appears despite the fact that, experimentally the pinch points become sharply developed at low temperature, indicating that the topological constraints are imposed to an excellent approximation\cite{Fennell2009}.

\begin{figure}[ht]
\includegraphics[trim=50 220 80 220,clip=true,scale=0.5]{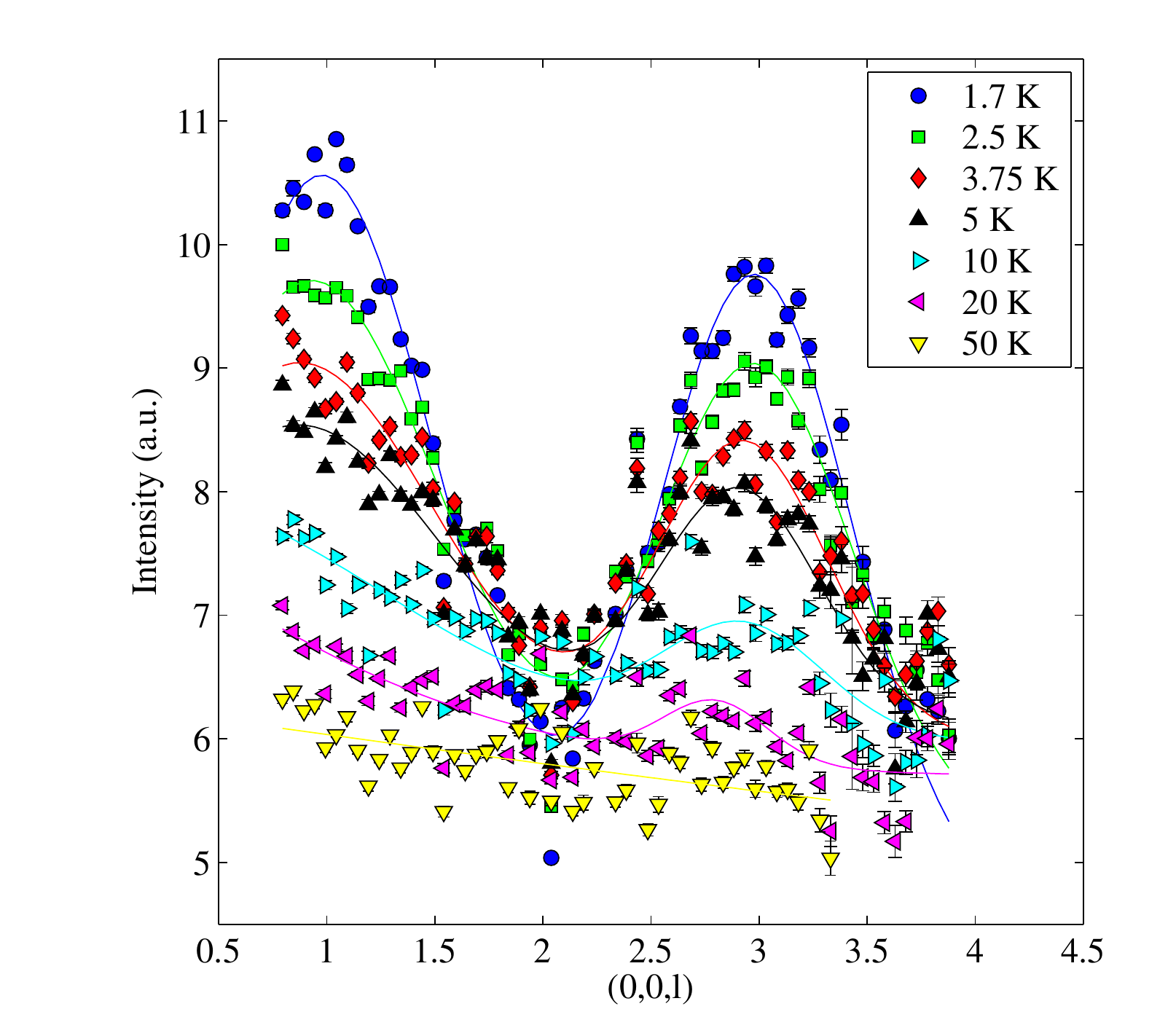}
\caption{Neutron scattering amplitude along the ridge $(0,0,l)$ taken from the second experiment~\cite{Fennell2009}, for temperatures between $50$ K and $1.7$ K. The lines are a guide to the eye.}
\label{fig:cuts}
\end{figure}

A more detailed look at the evolution of the scattering intensity along the ridge is shown in Fig. \ref{fig:cuts} where $S(\mathbf{Q})$ is plotted as a function of  $\mathbf{Q}(0,0,l\,2\pi/a)$ from $l\sim 0.7$ to $l=4$ and for temperatures between $50$ and $1.7$ K. This data is extracted from polarized diffuse scattering measurements made using the D7 spectrometer at the ILL, Grenoble, which are described in Ref~\cite{Fennell2009}. Again different polarization channels have been combined so that the total magnetic response is plotted.
 On the zone boundaries, with $l=1$ and $l=3$, the crossover between Curie laws is clearly visible, while at the zone centre, $l=2$ the crossover fails to develop. Please note that [1.7 K ; 50 K] is not a wide enough temperature window to see the full factor of 2 between the two Curie laws. The variation between both Brillouin zone boundaries ($l=1$ and $3$) is due to the $\mathbf{Q}$ dependence of the atomic form factor, which is not present in the simulations.
Without the form factor, the scattering amplitude would be equal for $l=1$ and $3$, in complete agreement with the numerical simulation.

This difference in behavior at the centre and zone boundaries shows that the Pauling states with strings spanning large distances are partially suppressed in the real material, so that the topological sector fluctuations, defined over large length scales, are less intense than those expected from a pure ice rule system in which all constrained microstates have the same statistical weight. Indeed, as discussed in Ref.~\cite{Yavors'kii2008} the scattering in the closely related spin ice material Dy$_2$Ti$_2$O$_7$ tends towards a pattern characteristic of closed 6-membered 
loops, though retains weak pinch points and the $00l$ ridge. A first possible explanation is suppression of long loops by further neighbour exchange couplings~\cite{Yavors'kii2008}; another is the closing of field lines - similar to strings - by the magnetic dipole interaction, while a third and more interesting possibility, is condensation into resonating loops stabilized by quantum fluctuations~\cite{Sikora2009,Sikora2011,Shannon2012}. Interestingly, the bulk susceptibility measurements from the powder sample are more in line with the topological sector fluctuations of the NNSI than the results for scattering at zero wave vector (Fig.~\ref{fig:SF:chiQ}). This could be due to a difference between results for single crystals and powder samples \cite{Hiroi2011} but it is an interesting point that requires further investigation.
In Fig. \ref{fig:SF:chiQ} we also show the evolution of the scattering intensity at points along the $(h,h,0)$ axis. The data for $h=3/2$ and $h=2$ become similar to the numerical scattering data from the NNSI  when it is scaled to the theoretical susceptibility at high temperature. As temperature is reduced, the intensity increases slightly before approaching a plateau, well below the theoretical predictions for scattering along the cubic axis [001]. This is to be expected as the spin components perpendicular to this low symmetry direction fail to capture the correlations of the Coulomb phase, as can be read from equation~(\ref{spinch}).

This data presentation allows us to make some comments regarding the relaxation of the system towards equilibrium, which  becomes very slow at low temperature ($\lesssim 0.6$ K), as indicated by field cooled, zero field cooled splitting of the magnetization~\cite{Snyder2004a}. This can be understood as a topological ergodicity breaking expected in classical systems~\cite{Castelnovo2007} where the rarefaction of topological defects, that can be seen as monopoles in presence of dipolar interactions~\cite{Ryzhkin2005,Castelnovo2008,Bramwell2009}, hinders the magnetic relaxation in both numerical simulations and in experiment~\cite{Matsuhira2000,Ehlers2003,Snyder2004a,Jaubert2009b}. However the unpolarized neutron scattering data points at $(0,0,0.9)$ measured on IN14 follow the analytical curve down to $\sim 0.1 - 0.2$ K, which suggests that equilibrium of large wave vector components of the magnetization can be achieved in spin ice in zero field down to temperatures significantly below the ergodicity temperature of 0.6 K. Less can be said about large length scales as loss of equilibrium does not appear to be the mechanism for the dip in scattering intensity along the $(0,0,l)$ ridge, at small wave vector. The deviation of $S(\mathbf{Q})$ from the theoretical prediction, at the zone centre occurs on the scale of $10$ K, well above the ergodicity temperature.

\paragraph{Topological sectors in other systems:}
Finally, we address the question of generality, as a
similar enhancement of paramagnetic fluctuations occurs in other topologically constrained systems if one has access to the appropriate variables. The enhanced fluctuations are a result of the emergent gauge phenomena \cite{Huse2003} that characterize the Coulomb phase in magnetic pyrochlore and kagom\'e systems \cite{Canals2001,Isakov2004,Henley2005,Conlon2010,Macdonald2011}. The emergent field is subject to a local divergence free constraint, leading to dipolar correlations and the structure of topological sectors described above. 
In spin ice models and materials the emergent gauge field is proportional to the coarse grained field of magnetic moments, or spin configuration. Hence it is directly accessible through both bulk measurement and  through scattering experiments. The direct access to the gauge field~\cite{Fennell2009} makes spin ice of particular interest in this context, as we have shown in the present paper. 
In pyrochlore antiferromagnets the emergent gauge field is a hidden property of the rule of satisfied units\cite{Moessner2001} and essentially corresponds to an inverse mapping between antiferromagnetically coupled spins and ferromagnetically coupled pseudo-spins that resemble the spin ice degrees of freedom and form an effective magnetic field within a coarse grained description~\cite{Isakov2004}.
Hence the physics of the emergent gauge field is common to both classes of system. This is the case for the Heisenberg antiferromagnet:
within a Gaussian approximation, the susceptibility for the pseudo degrees of freedom is enhanced by the same factor of two as the susceptibility in spin ice, as one descends into the Coulomb phase \cite{Conlon2010,Benton2012}. The pseudo spin degrees of freedom are not accessible through bulk measurement, but magnetic scattering in $\mathbf{Q}$ space is related to that for spin ice, with the topological constraints leading to a pinch point pattern. For nearest neighbour interactions only, one finds lines of constant scattering amplitude for the real spins, this time along the $(h,h,0)$ axes. As further neighbour interactions are added, the intensity dips at the Brillouin zone centre, even though the lowest energy states remain within the topologically constrained phase space. This result seems in complete analogy with our experimental observations along the $(0,0,l)$ axis in spin ice, giving weight to our interpretation in terms of topological sector fluctuations of varying intensity, as the effective length scale, $\ell \sim 1/|\mathbf{Q}|$ changes from the zone centre to zone boundary.

The factor of two of this crossover is directly related to the entropic weight of the different topological sectors in the Coulomb phase~\cite{Conlon2010,Benton2012}. Hence it is \textit{not} universal and we can expect the \textit{spin liquid Curie} laws in other frustrated systems to have a different prefactor.

\paragraph{Conclusion:} The local divergence free constraint of the Coulomb phase in spin ice materials allows a decomposition of the ground state ensemble in terms of topological sectors. Fluctuations between sectors is clearly visible through susceptibility measurements, as the winding numbers characterizing the sectors are directly proportional to the magnetization. This is only true in the Coulomb phase and as the system settles into it, dipolar spin correlations develop, giving  a \textit{Curie law crossover} between paramagnetism and fluctuations characteristic of the topological spin liquid phase. We have compared analytical and numerical results for the nearest neighbour spin ice model with bulk magnetometry measurements for a powder and neutron scattering measurements through the structure function $S(\mathbf{Q},T)$ for a single crystal of \hto{}. We find quantitative agreement between theory and experiment for large $\mathbf{Q}$, but near the pinch point, at small wave vector the experimental scattering intensity is suppressed compared with theory. We have discussed how further nearest neighbour exchange and dipole interactions, or quantum fluctuation, may hinder topological sector fluctuations on a large length scale. Because such perturbations can be pertinent in real materials, especially dipolar interactions for spin ice compounds, the success of the nearest neighbour model raises interesting open questions. More experimental and theoretical work is required to understand these mechanisms in details.

We believe this Curie law crossover is in fact a general feature of many frustrated systems~\cite{Jaubert}, apparent as TSF in spin ice materials and encoded into the scattering pattern of related antiferromagnets. As a consequence, the standard Curie-Weiss picture at high temperature appears to be incomplete and should be used with caution. Further work in this direction, and in particular in potential quantum spin liquid compounds - Herbertsmithite~\cite{Mendels2007}, Tb$_{2}$Ti$_{2}$O$_{7}$~\cite{Gardner2003,Gardner2010}, Yb$_{2}$Ti$_{2}$O$_{7}$~\cite{Ross2011,Thompson2011} - would be particularly interesting.

It is remarkable that, in spin ice, a completely local probe readily accessible by experiments is able to identify fluctuations between topological sectors; i.e. the difference between ``topological constraints'' and ``topological order''.  Until now, this task necessarily fell to non-local probes, such as measures of the winding number, or the topological entanglement entropy~\cite{Kitaev2006,Levin2006}. In gapless U(1) quantum liquids, the notion of topological entanglement entropy is expected to be ill-defined~\cite{Ju2012}. In light of the revelation that there may be other, yet undiscovered topological invariants in related systems \cite{Ran2011,Freedman2011}, it is interesting to speculate whether similar probes may prove instrumental in characterizing this important class of topological order in the future.

\paragraph{Acknowledgements} It is a pleasure to thank  A. Harman-Clarke, M. Gingras, M. Hastings, L.-P. Henry, G. Misguich, R. Moessner, T. Roscilde and N. Shannon for useful discussions; A.R. Wildes and M. Boehm for contributions to the second IN14 experiments; O. Losserand  for cryogenic support during the second experiment and D. Prabhakaran for a sample. We are grateful for financial support from the European Science Foundation PESC/RNP/HFM (LDCJ and PCWH), the Ecole Normale SupŽrieure de Lyon (RGM) and the Institut Universitaire de France (PCWH).

\bibliographystyle{ludo_aps}
\bibliography{library}

\end{document}